\documentclass[floatfix,showpacs,superscriptaddress,prl,twocolumn]{revtex4}
\usepackage{amsbsy,amssymb,amsmath,bm}
\usepackage{epsf}
\usepackage{graphicx,color}
\usepackage{amsfonts}
\usepackage{amssymb}
\usepackage{amsmath}

\definecolor{red}{rgb}{1,0,0}           
\definecolor{green}{rgb}{0,1,0}
\definecolor{blue}{rgb}{0,0,1}
\definecolor{darkblue}{rgb}{0,0,0.5}
\definecolor{lightblue}{rgb}{.5,.5,1}
\definecolor{lightgray}{gray}{.87}          
\definecolor{Dark}{gray}{.20}
\definecolor{pink}{rgb}{.95,0.82,0.92}  
\definecolor{yellow}{rgb}{1,1,0}
\definecolor{lightyellow}{rgb}{1,1,.5}
\definecolor{purple}{rgb}{0.7,0,0.85}
\definecolor{darkgreen}{rgb}{0,0.5,0}
\definecolor{orange}{rgb}{0.8,0.2,0.2}

\def \be {\bea}
\def \ee {\eea}
\def \bea {\begin{eqnarray}}
\def \eea {\end{eqnarray}}
\def \bse {\begin{subequations}}
\def \ese {\end{subequations}}
\def \bde {\begin{description}}
\def \ede {\end{description}}
\def \nn {\nonumber}

\def \dels {\partial\kern-.5em / \kern.5em}
\def \As {{A\kern-.5em / \kern.5em}}
\def \Ds {D\kern-.7em / \kern.5em}

\def \b {\beta}
\def \dag {\dagger}

\def \d {\delta}
\def \D {\Delta}

\def \m {\mu}

\def \lam {\lambda}

\def \s {\sigma}

\def \t {\tau}

\def\Id{\mathbb{I}}
\def\Mp{\mathbb{M}_+}
\def\Mm{\mathbb{M}_-}

\def \mef {\mu_{\rm eff}}
\def \HH {{\mathbb{H}}}
\def \UU {{\mathbb{U}}}

\def \SS {{\mathbb{S}}}

\begin{document}

\title{Chiral zero modes in superconducting nanowires with Dresselhaus spin-orbit coupling}
\author{Hsien-chung Kao}
\affiliation{\textit{Physics Department, National Taiwan Normal University, Taipei 11677,
Taiwan}}
\email{hckao@ntnu.edu.tw}
\affiliation{\textit{Physics Department, Simon Fraser University, Burnaby, BC Canada}}
\date{\today }

\begin{abstract}
Using chiral decomposition, we are able to find analytically the zero modes and the conditions for such modes to exist in the Kitaev ladder model and superconducting nanowires with Dresselhaus spin-orbit coupling.  As a result, we are able to calculate the number of zero modes in these systems for arbitrary given parameters in the semi-infinite limit.  Moreover, we find that when suitable resonance condition is satisfied exact zero modes exist even in finite systems contrary to the common belief. 
\end{abstract}

\pacs{74.20.-z, 74.78.-w, 74.25.F-,71.10.Pm}
\maketitle

\subsection{I. Introduction}

In recent years, the Majorana fermion has been a subject under intensive investigation in condensed matter physics \cite{Alicea, Beenakker1} . It usually emerges as Majorana zero energy states (MZES) in systems exhibiting topologically non-trivial phase. Typically MZES appear in pairs, clinging to the edges separated by a macroscopic distance, which makes them nonlocal. In addition, systems with many pairs of MZES may obey nonabelian statistics.  Therefore, it is believed that MZES may be used to realize fault-tolerant quantum bits, and thus would play an important role in quantum computation \cite{Ivanov,Sau1}. The Kitaev model is probably the simplest one that support MZES \cite{Kitaev}. In addition to the typical nearest-neighbor hopping term, there is also a $p$-wave superconducting amplitude in the model. When the chemical potential of the system is in proper range, MZES show up as edge bound states, whose existence signifies that the system is in the topologically non-trivial phase.  They are also responsible for the zero-bias conductance of the system. Since the Kitaev model was proposed, there have been many extensions by various authors \cite{Potter,Niu,Rieder,Kells,Asahi,Zhou,Manmana,DeGottardi1,Wimmer,Gangadharaiah,Nagaosa}. There are also many other systems that give rise to MZES\cite{Fu,Oreg,Lutchyn1}.  

To observe MZES in experiments, it has been proposed to take advantage of the proximity effect and put semiconductor nanowires on top of a $s$-wave superconductor. To mimic the effect of a $p$-wave superconductor, one must use a semiconductor with strong spin-orbit interaction and introduce a Zeeman field. With a Rashba spin-orbit coupling, such a system may have a topologically no-trivial phase \cite{Oreg,Lutchyn1,Stanescu}.  However, it is difficult to show a compelling evidence experimentally, since it is a delicate matter to tune the number of propagating channels in the nanowires without spoiling the superconductivity \cite{Potter,Oreg,Lutchyn1,Lutchyn2,Gibertini}. Another possibility is to utilize semiconductors with a strong Dresselhaus spin-orbit coupling, such as InSb or GaAs \cite{Dresselhaus}. When an in-plane Zeeman field is applied, such systems may achieve topologically non-trivial phase.  As the system belongs to the BDI class, its number of MZES may take on arbitrary integer values\cite{Sato1,Ikegaya}. Although some signature of MZES has been reported in experiments \cite{Mourik,Das,Deng,Rokhinson}, it is still under debate whether the zero-bias anomaly really stems from Majorana fermions \cite{EJHLee1,EJHLee2,Liu,Rainis}. 

Unfortunately, it is usually quite difficult to find analytic solutions in most of the systems we are interested in. We notice that such a situation may be improved when there is a chiral operator in the system.  Since such an operator anti-commutes with the Hamiltonian of the system, energy eigenstates always appear in pairs.  In particular, the zero energy states can be divided into decoupled left and right-handed chirality states.  As a result, the Bogoliubov de-Gennes equation in the zero energy subspace is greatly simplified and can be exactly solved.  In this paper, we will explicitly show how to exploit the chiral decomposition to carry out analytic study of the MZES in the Kitaev ladder model and superconducting nanowires with the Dresselhaus spin-orbit coupling.   

The rest of the paper is organized in the following way. In Sec. II, we first apply the chiral decomposition to the Kitaev model as an illustration of the method. By decoupling the left and right-handed chirality zero energy states, we are able to solve the equation analytically. By taking advantage of the explicit form of the solutions, we find that exact zero energy states exist even for a finite system, when certain resonance condition is met.  This can be done by fine tuning of the chemical potential. In Sec. III, we then carry out similar analysis in the Kitaev ladder model in which the superconducting order parameter has the same phase in the $x$ and $y$- directions so that it has a chiral operator. Again, we are able to find the analytic solution of the chiral zero energy states. When resonance condition is achieved, exact zero energy states also exists even for finite systems.  By making use of the analytic form of the condition, we are able to calculate the number of MZES for arbitrary given parameters.  In Sec. IV, we proceed to apply similar analysis to superconducting nanowires with the Dresselhaus spin-orbit coupling. We observe that the $y$-direction maybe be diagonalized easily, and the system effectively becomes one-dimensional. When the system is semi-infinite, we are able to derive the exact condition for the existence of zero energy states, which may be used to find out the number of zero modes analytically in the whole parameter space.  Here, exact zero-energy states also exist for a finite system. Although the resonance condition is complicated in terms of the physical parameters, we may still use numerical calculation to establish that such states do exist. Finally, we make conclusion and discuss possible extensions in Sec. V.

\subsection{II. Chiral decomposition of the Kitaev chain}

Let's begin with the well-known simple Kitaev chain to show how chiral decomposition works. The Hamiltonian is given by
\bea
&\;& H_{1K}=\sum_{j=1}^{N}\left\{ -\m c_j^\dag c_j \right\} + \sum_{j=1}^{N-1}\left\{ -w c_j^\dag c_{j+1} + {\rm H.c.} \right\} \cr
&\;& \hskip 0.75cm + \sum_{j=1}^{N-1}\left\{ -\D\, c_j^\dag c_{j+1}^\dag + {\rm H.c.} \right\}.
\eea
Here, $j$ indicates the lattice coordinate, $\m$ the chemical potential, $w$ the hopping parameter, and $\D$ the superconducting pairing\email{hckao@ntnu.edu.tw} amplitude.  Without loss of generality, we will assume $w$ to be positive through out the paper for convenience. The Bogoliubov de-Gennes Hamiltonian is given by 
\bea
&\;& \HH_{1K}=\t_3\otimes\left(-\m \Id - w \Mp \right) + \t_2\otimes\left(-i\D \Mm \right).
\eea
Here,  $\t_1, \t_2,\t_3$ are the Pauli matrices and $\Id_{ij}=\d_{i,j}, (\Mp)_{ij}=\d_{i+1,j}+ \d_{i,j+1},$ $(\Mm)_{ij} = \d_{i+1,j}- \d_{i,j+1},$ being $N\times N$ matrices. 
It is obvious that $\HH_{1K}$ anti-commutes with the chiral operator $\Pi = \t_1$, which means that eigenstates of $\HH_{1K}$ generally appear in pairs with eigenvalues $(E, -E)$ and wavefunction $(\UU, \t_1 \UU)$.  Because of this property, we can decompose the zero modes into left and right-handed chirality states, 
\bea
\UU_{\rm R} \equiv \frac{\Id+ \t_1}{2}\UU,\; \UU_{\rm L} \equiv \frac{\Id- \t_1}{2}\UU,
\eea
which would be both energy and chirality eigenstates. To be more specific, we may explicitly use the unitary transformation 
\be
\SS_1 = \frac{1}{\sqrt{2}} 
\left(\begin{matrix}
1 & -1\cr
1 & 1 \cr
\end{matrix}\right) \otimes \Id_x,
\ee
to bring $\HH_{1K}$ into the following form
\bea
&\;& \hskip -1.0cm \tilde{\HH}_{1K}= \t_+\otimes\tilde{\HH}_{1K,{\rm R}} + \t_-\otimes\tilde{\HH}_{1K,{\rm L}}, \cr
&\;& \hskip -2.1cm {\rm where}\; \t_\pm = \frac{1}{2}(\t_1 \pm i\t_2), \nn\cr
&\;& \hskip -1.7cm {\rm and}\; \tilde{\HH}_{1K,{\rm R}} = -\m \Id  -w \Mp - \D \Mm, \nn\cr
&\;& \hskip -1.0cm \tilde{\HH}_{1K,{\rm L}} = -\m \Id  -w \Mp + \D \Mm. \nn
\eea
As a result, $\tilde{\UU}= (\tilde{\UU}_{\rm L}, \tilde{\UU}_{\rm R})^T$ and the Bogoliubov de-Gennes equation becomes
\bea
&\;& \hskip -3.1cm \tilde{\HH}_{1K,{\rm R}} \tilde{\UU}_{\rm R} = E \tilde{\UU}_{\rm L}, \cr
&\;& \hskip -3.1cm \tilde{\HH}_{1K,{\rm L}} \tilde{\UU}_{\rm L} = E \tilde{\UU}_{\rm R}. 
\eea
In particular, the left and right-handed zero modes satisfy the following decoupled equations.
\bea
&\;& \hskip -2.0cm \tilde{\HH}_{1K,{\rm R}} \tilde{\UU}_{\rm R}=0, \quad \tilde{\HH}_{1K,{\rm L}} \tilde{\UU}_{\rm L}=0,
\eea
respectively. Since these are nothing but constant coefficient 2nd order difference equations, it is now simple to find the analytic solutions.  Let $[\tilde{\UU}_{\rm R}]_j = \tilde{u}_{0,{\rm R}}s^j$, and we see $s$ satisfies the following secular equation:
\be
f(s)=-\m -(w+\D)s-(w-\D)s^{-1}=0.
\ee
The solutions are 
\be
s_\pm=\frac{-\mu\pm\sqrt{\m^2-4(w^2-\D^2)}}{2(w+\D)}.
\ee
As there are two solutions to the above equation, the general form of $\tilde{u}_R(j)$ is given by
\be
\tilde{u}_{\rm R}(j)=a_+ s_+^j+a_- s_-^j.
\ee
After simplification, the equation of motion at site $j=1,N$ leads to the boundary condition 
\bea
\tilde{u}_{\rm R}(0)=0,\; {\rm and}\;\; \tilde{u}_{\rm R}(N+1)=0,
\eea 
respectively. Therefore, we have
\bea
&\;& a_+ + a_- =0, \cr
&\;& a_+ s_+^{N+1} + a_- s_-^{N+1}=0.
\eea
They can be satisfied only if the solution of $\tilde{u}_R(j)$ is oscillatory, i.e. $w>|\D|$, and $|\m|<2\sqrt{w^2-\D^2}$. 
For such a case, we have 
\bea
&\;& \m=2\sqrt{w^2-\D^2}\cos p \cr
&\;& s_\pm= -\sqrt{\frac{w-\D}{w+\D}} e^{\mp i p}, \cr
&\;& \hskip -0.8cm {\rm with}\; p = \frac{\pi k}{N+1}, \; k=1,\ldots, N. \label{zero_K1}
\eea
Similar results hold for the left-handed modes, where
\bea
&\;& s_\pm= -\sqrt{\frac{w+\D}{w-\D}} e^{\mp i p}, \cr
&\;& \hskip -0.8cm {\rm with}\; p = \frac{\pi k}{N+1}, \; k=1,\ldots, N. 
\eea
Given $w$ and $\D$, we therefore always have a pair of exact zero modes for $N$ discrete values of $\m$. This is different from the common belief that exact zero modes only exist in semi-infinite systems.  We show a numerical check for the above analytic result (Fig.1). In all the numerical results in this paper, we use $w$ as the unit of the energy. It is clear from the figure that exact zero modes for a finite system only exist for $w>|\D|.$    
\begin{figure}[t]
\centerline{\includegraphics[width=0.45\textwidth]{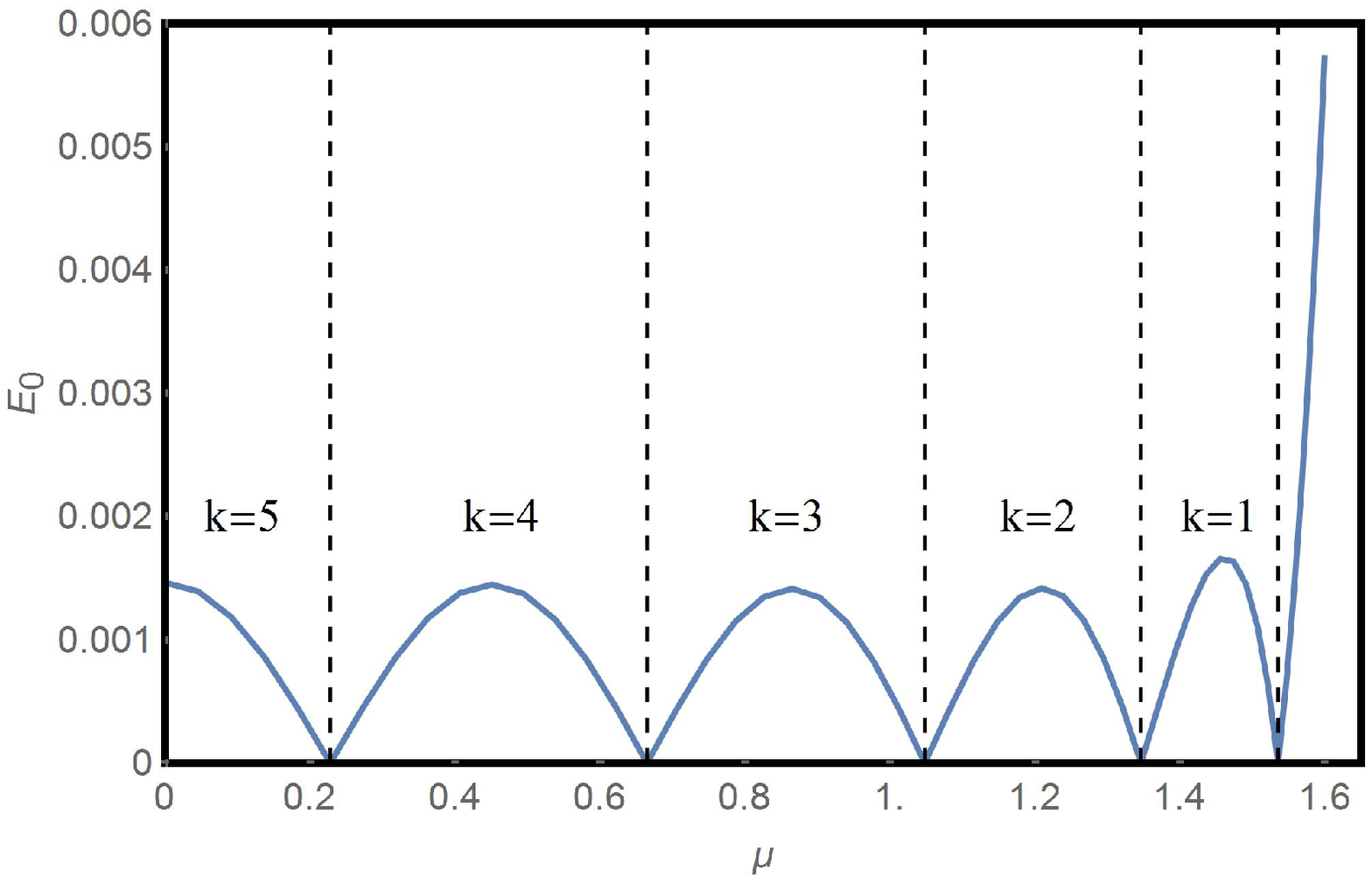}}
\centerline{\includegraphics[width=0.45\textwidth]{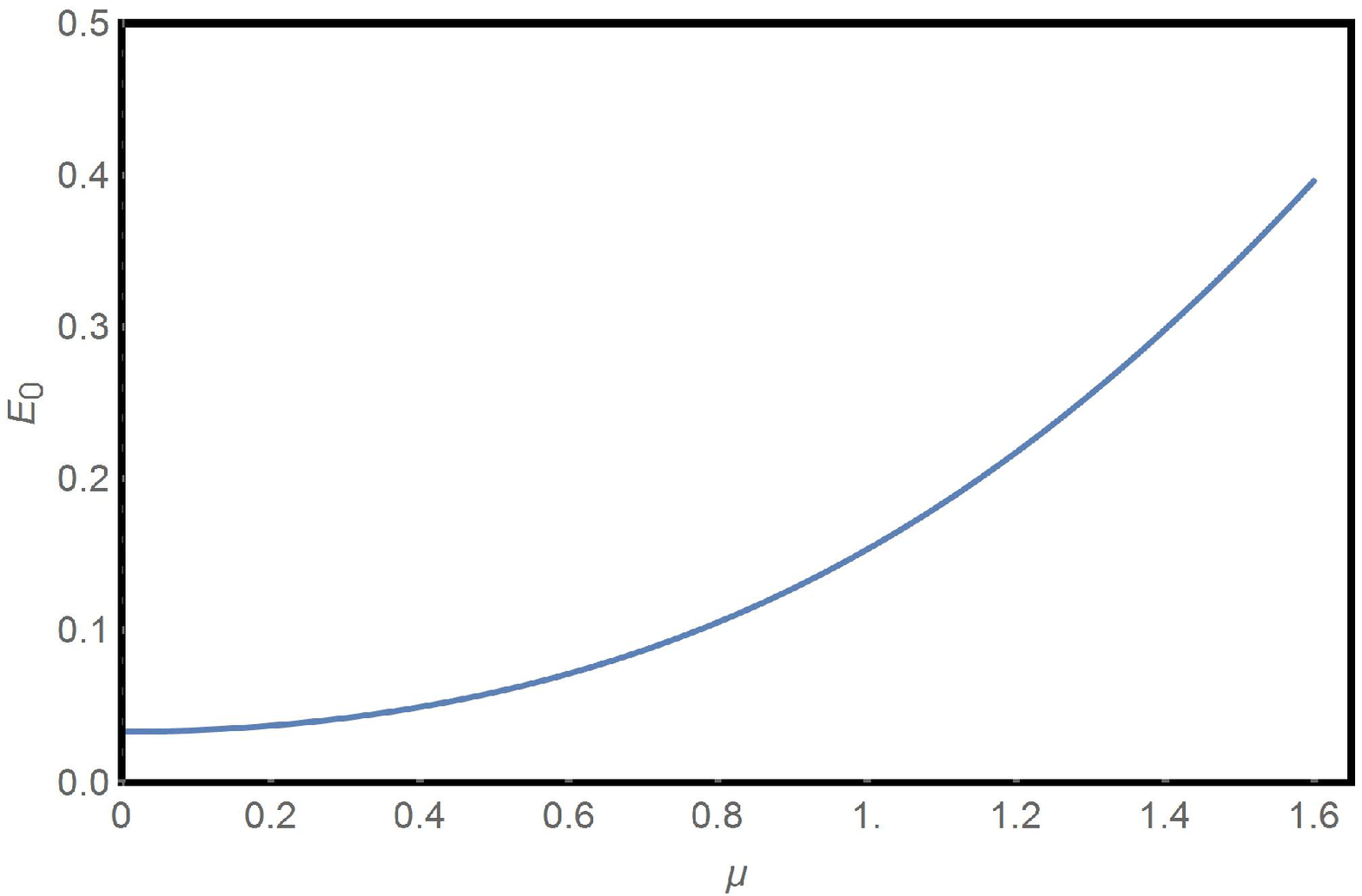}}
\caption{$E_0$  vs. chemical potential $\m$ for the Kitaev chain is shown, where $E_0$ is the energy of the lowest positive energy mode.  We choose $N=10$ so that the energy difference between the exact zero modes and would-be zero modes is more manifest. (a) top. When $w>|\D|$,(numerically $\D=0.6$ here), there are a pair of exact zero modes when the condition in Eq.(\ref{zero_K1}) is satisfied. The value of the exact zero mode energy is zero up to numerical inaccuracy, which is of the order $10^{-16}.$ (b) bottom. When $w<|\D|$, there is no zero mode even if $|\m|<2w.$ Here, we choose $\D=2.4$ so that the effect is more pronounced. }
\label{E0_vs_mu_Kitaev_1d}
\end{figure}

If we take the $N\to\infty$ limit of the above results, both the left and right-handed zero modes will seemingly have a continuous spectrum.  However, since the left and right-handed solutions to its corresponding secular equation always form reciprocal pairs, only one of them would survive the infinite $N$ limit. More specifically, it is the right-handed (left-handed) modes that will survive for $\D>0$ $(\D<0)$.  Thus, the limit is not a smooth one.  For definiteness, let us assume $\D>0$ here.  The boundary condition at $j=0$ still imposes $a_+ = -a_- =a_0$ on the right-handed modes, and we have 
\be
\tilde{u}_{\rm R}(j)=a_0(s_+^j - s_-^j).
\ee
In contrast to the finite $N$ case, the boundary condition at infinity can always be satisfied if we have both $|s_+|, |s_-| < 1$.  This translates to the condition that
\bea
-2w<\m<2w,
\eea
and the condition $w>|\D|$ is lifted.  It can be seen that the above condition is equivalent to 
\bea
f(1)f(-1)<0, 
\eea
which is exactly the Paffian of the zero modes, as is well-known in the literature \cite{Wen,DeGottardi2}.  Since $f(s)$ diverges at $s=0$, the mathematical implication of the above condition is that the secular equation has either no or two real roots in the interval $[-1, 1]$. 

The fact that only one of the left and right-handed solutions can survive in the infinite $N$ limit always holds for the models we consider, since their solutions to the corresponding secular equation always form a reciprocal pair, as we will see later.  This suggests that chiral symmetry of the zero modes is broken in a semi-infinite system. It would be interesting to find out what this implies in the context of lattice field theories.

\subsection{III. Chiral decomposition of the Kitaev Ladder Model}
To show the powerfulness of the chiral decomposition method, we now proceed to carry out similar analysis for the two-dimensional generalization of the Kitaev chain.  The Bogoliubov de-Gennes Hamiltonian of the Kitaev ladder model takes the form \cite{Nagaosa}
\bea
&\;& \hskip -0.6cm \HH_{2K}=\t_3\otimes\left\{-\m \Id_x\otimes \Id_y -w_x (\Mp)_x \otimes \Id_y - w_y \Id_x \otimes (\Mp)_y  \right\} \nn \cr
&\;& \hskip  0.3cm + \t_2\otimes\left\{-i\D_x (\Mm)_x \otimes \Id_y -i\D_y \Id_x \otimes (\Mm)_y \right\}.
\eea
Here, we may as well assume both $w_x$ and $w_y$ to be positive without loss of generality. For the system to have a chiral operator which anti-commute with the Hamiltonian, we will only consider the case that there is no relative phase between $\D_x$ and $\D_y$ so that we may make both of them real with suitable phase rotation.  Again, we may decompose the Hamiltonian into the left and right-handed parts using a similar unitary transformation:
\bea
&\;& \tilde{\HH}_{2K} = \t_+\otimes\tilde{\HH}_{2K,{\rm R}} + \t_-\otimes\tilde{\HH}_{2K,{\rm L}}, \cr
&\;& \hskip -0.8cm {\rm with}\; \tilde{\HH}_{2K,{\rm R}} = -\m \Id_x\otimes \Id_y - w_x (\Mp)_x\otimes \Id_y - w_y \Id_x\otimes(\Mp)_y  \nn\cr
&\;& \hskip 1.4cm - \D_x (\Mm)_x\otimes\Id_y - \D_y \Id_x\otimes(\Mm)_y, \nn\cr
&\;& \tilde{\HH}_{2K,{\rm L}} = -\m \Id_x\otimes \Id_y - w_x (\Mp)_x\otimes \Id_y - w_y \Id_x\otimes(\Mp)_y  \nn\cr
&\;& \hskip 1.4cm + \D_x (\Mm)_x\otimes\Id_y + \D_y \Id_x\otimes(\Mm)_y. \nn
\eea
Now, the Bogoliubov de-Gennes equation becomes
\bea
&\;& \hskip -3.1cm \tilde{\HH}_{2K,{\rm R}} \tilde{\UU}_{\rm R}(j_x, j_y) = E \tilde{\UU}_{\rm L}(j_x,j_y), \cr
&\;& \hskip -3.1cm \tilde{\HH}_{2K,{\rm L}} \tilde{\UU}_{\rm L}(j_x,j_y) = E \tilde{\UU}_{\rm R}(j_x,j_y).
\eea
Again the left and right-handed zero modes are decoupled and they satisfy 
\bea
&\;& \hskip -3.1cm \tilde{\HH}_{2K,{\rm R}} \tilde{\UU}_{\rm R}=0, \quad \tilde{\HH}_{2K,{\rm L}} \tilde{\UU}_{\rm L}=0,
\eea
respectively. 

The analytic solutions may be found in a similar way.
Let $[\tilde{\UU}_{\rm R}]_{j_x,j_y}= \tilde{u}_{0,{\rm R}}s_x^{j_x} s_y^{j_y}$, with $s_s, s_y$ satisfying the following secular equation:
\bea
&\;& \hskip -2.6cm -\m -(w_x+\D_x)s_x -(w_x-\D_x)s_x^{-1} \cr
&\;& \hskip -2.0cm - (w_y+\D_y)s_y - (w_y-\D_y)s_y^{-1} =0.
\eea
After separation of variables, we have
\bea
&\;& \hskip -0.5cm f_x(s_x)=-\m_x -(w_x+\D_x)s_x-(w_x-\D_x)s_x^{-1} =0, \cr
&\;& \hskip -0.5cm f_y(s_y)=-\m_y -(w_y+\D_y)s_y-(w_y-\D_y)s_y^{-1} =0, \cr
&\;& \hskip -1.3cm {\rm with}\; \m = \m_x +\m_y.
\eea
The solutions are 
\bea
&\;& (s_x)_\pm=\frac{-\mu_x\pm\sqrt{\m_x^2-4(w_x^2-\D_x^2)}}{2(w_x+\D_x)}; \cr
&\;& (s_y)_\pm=\frac{-\mu_y\pm\sqrt{\m_y^2-4(w_y^2-\D_y^2)}}{2(w_y+\D_y)}.
\eea
For finite $N_x$ and $N_y,$ a pair of zero modes exist when $w_x>|\D_y|, w_y>|\D_y|$ and 
\bea
&\;& \hskip -0.5cm \m = 2\sqrt{w_x^2 -\D_x^2} \cos p_x + 2\sqrt{w_y^2 -\D_y^2} \cos p_y.
\eea
Here,
\bea 
&\;& \hskip -3.1cm  p_x = \frac{\pi k_x}{N_x+1},\; p_y = \frac{\pi k_y}{N_y+1}, \label{zero_K2_1}
\eea
with $k_x=1,\ldots N_x,\; k_y=1,\ldots, N_y.$

For the isotropic case, where $w_x=w_y=w, \D_x = \D_y=\D$ and $N_x=N_y=N$, more zero modes may occur simultaneously due to higher symmetry. In particular, there are $N$ pairs of zero modes for $\m = 0$ with $k_x,k_y$ in the required range and
\bea
&\;& \hskip -2.4cm k_x + k_y = N+1.
\eea
In contrast, there are two pairs of zero modes for  
\bea
&\;& \hskip -1.4cm k_x\neq k_y,\; {\rm and}\; k_x+k_y \neq N+1; 
\eea
one pair of zero modes for  
\bea
&\;& \hskip -1.4cm k_x=k_y,\; {\rm and}\; k_x+k_y \neq N+1. 
\eea
We show a numerical check for the isotropic case of the above analytic results in Fig.2. It can be seen clearly exact zero modes for a finite system only exist for $w>|\D|,$ and there may be more than one pairs of zero modes for some values of $\m$.    
\begin{figure}[t]
\centerline{\includegraphics[width=0.45\textwidth]{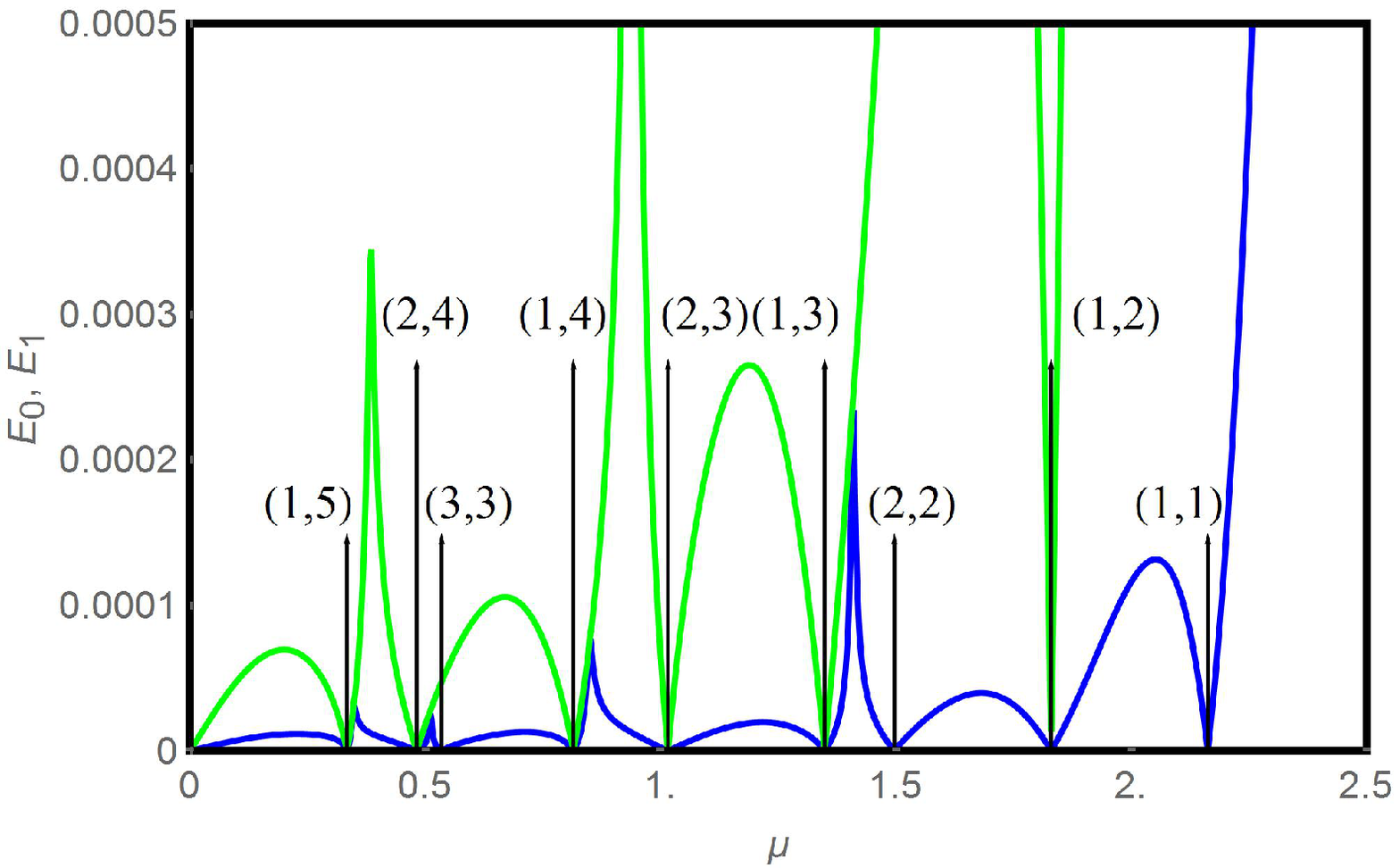}}
\centerline{\includegraphics[width=0.45\textwidth]{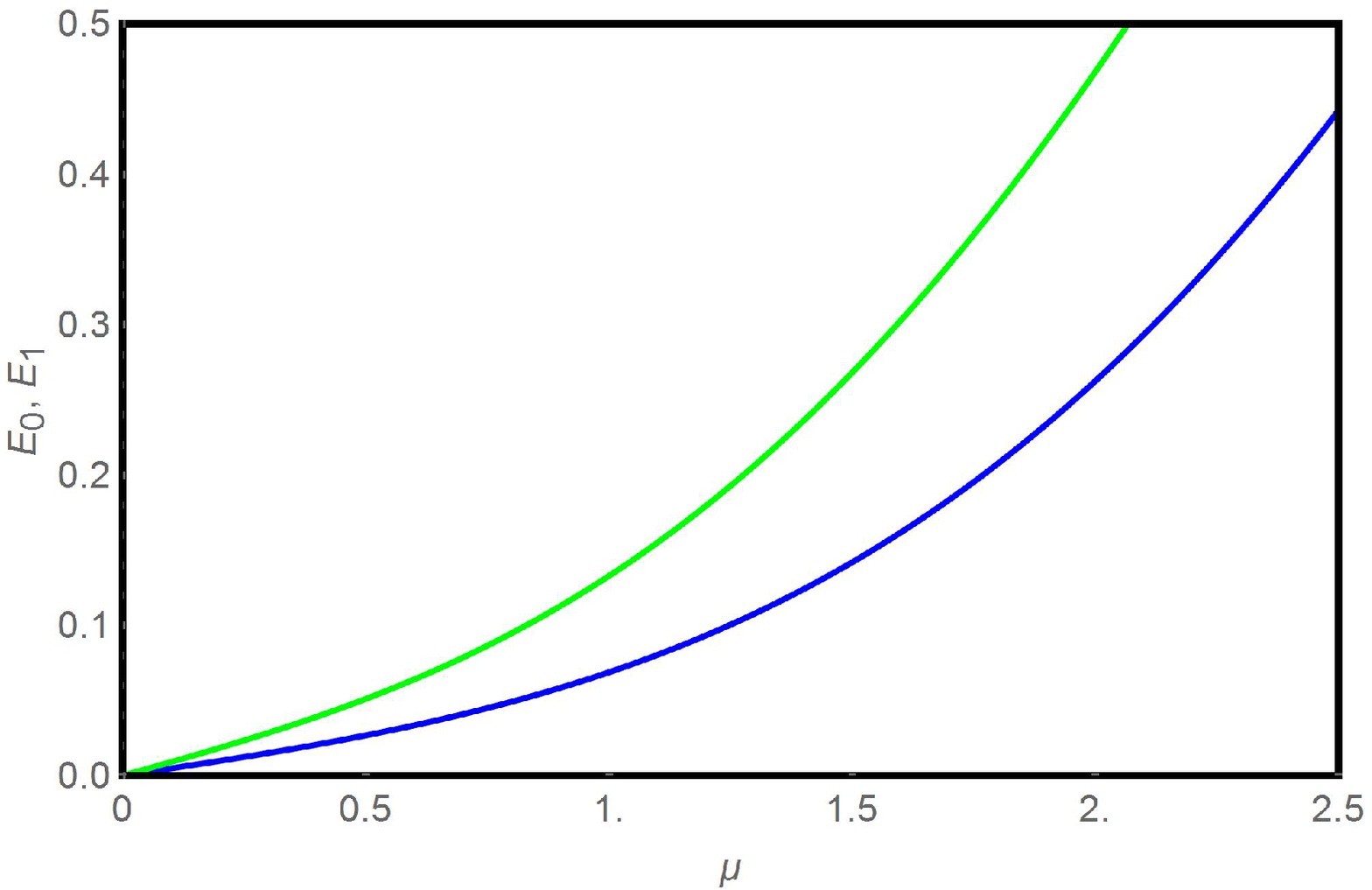}}
\caption{$E_0, E_1$ vs. chemical potential $\m$ for the Kitaev ladder models with $N_x = N_y =6$ is shown, where $E_0, E_1$ are the energies of the two lowest positive energy modes.  (a) top. For $w>|\D|$,(numerically $\D=0.8$ here), there will be exact zero modes when the condition in Eq.(\ref{zero_K2_1}) is satisfied. The two numbers in the parenthesis are the corresponding $k_x$ $k_y$ for the mode, and there are more than one pair of exact zero modes in some specific values of $\m$ as predicted in Eq. (24) and (25). Again, the value of the exact zero mode energy is zero up to numerical inaccuracy, which is of the order $10^{-16}.$ (b) bottom. For $w<|\D|$, there is no zero mode even if $|\m|<4w.$ Here, we choose $\D=2.4$ so that the effect is more pronounced. }
\label{E0_vs_mu_Kitaev_2d_1}
\end{figure}

Let us return to the general anisotropic case and take $N_x$ to be infinite.  As a result, the condition on the zero modes in the $x$-direction is lifted.  Hence, we have
\bea
&\;& \hskip -2.0cm -2w_x < \m_x < 2w_x. \cr
&\;& \hskip -2.0cm w_y>|\D_y|, \;{\rm and}\; \m_y =2\sqrt{w_y^2-\D_y^2}\cos p_y.
\eea
Whether the left or right-handed modes survive depends on the sign of $\D_x.$
By combining the above conditions, the following result may be obtained 
\bea
&\;& \hskip -2.0cm  \frac{\m -2w_x}{2\sqrt{w_y^2-\D_y^2}} <\cos\left( \frac{\pi k_y}{N_y+1} \right) <\frac{\m +2w_x}{2\sqrt{w_y^2-\D_y^2}}. \label{zero_K2_2}
\eea 
This may be used to determine the number of zero modes of the system given $\m, w_x, w_y, \D_y$ and $N_y$, which may be compared to previous numerical results in the literature \cite{Nagaosa}.  We show a numerical check of the above analytic results of a semi-infinite Kitaev ladder in Fig.3. It can be seen that there is no zero mode if $w<|\D|$.

\begin{figure}[t]
\centerline{\includegraphics[width=0.45\textwidth]{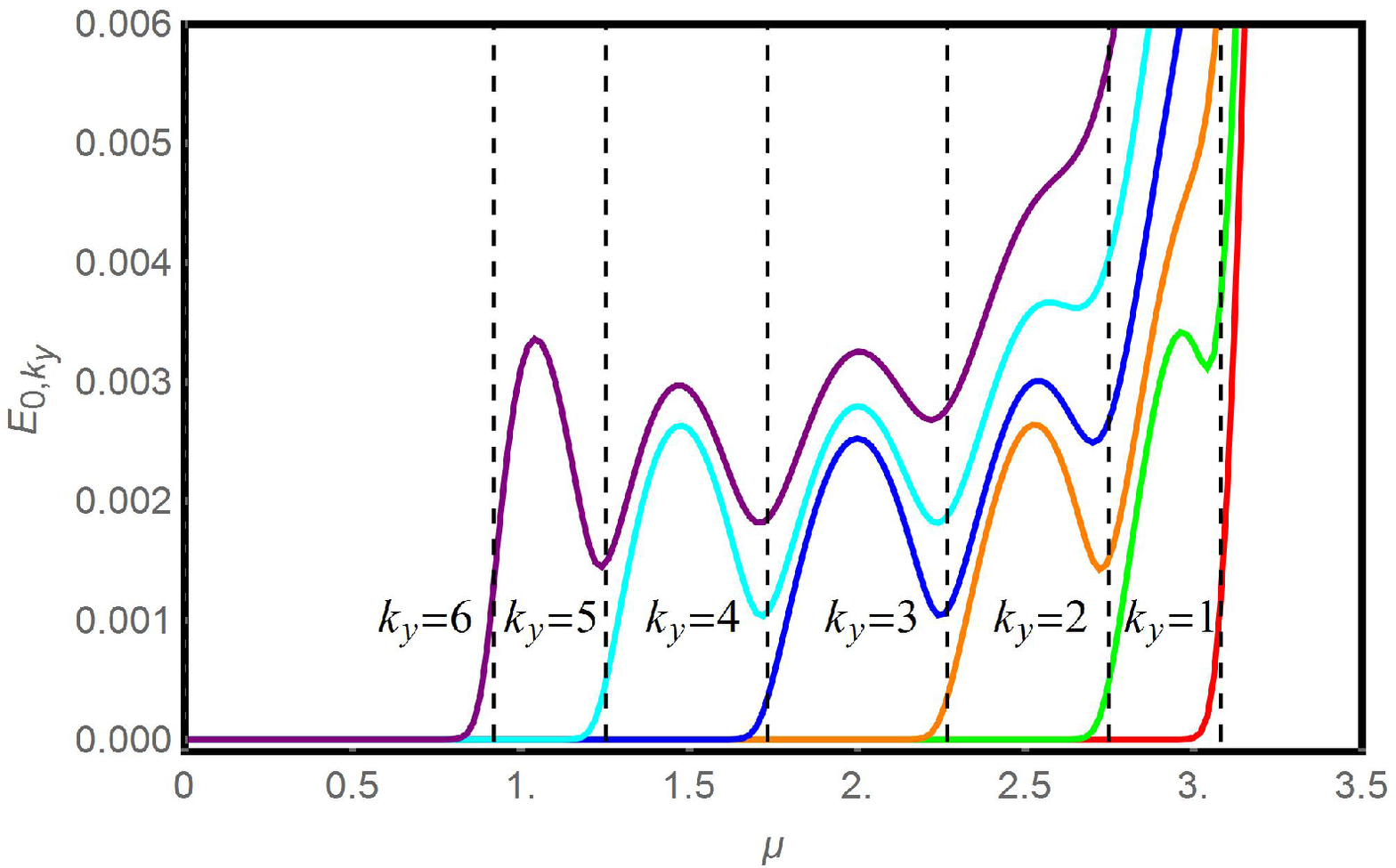}}
\centerline{\includegraphics[width=0.45\textwidth]{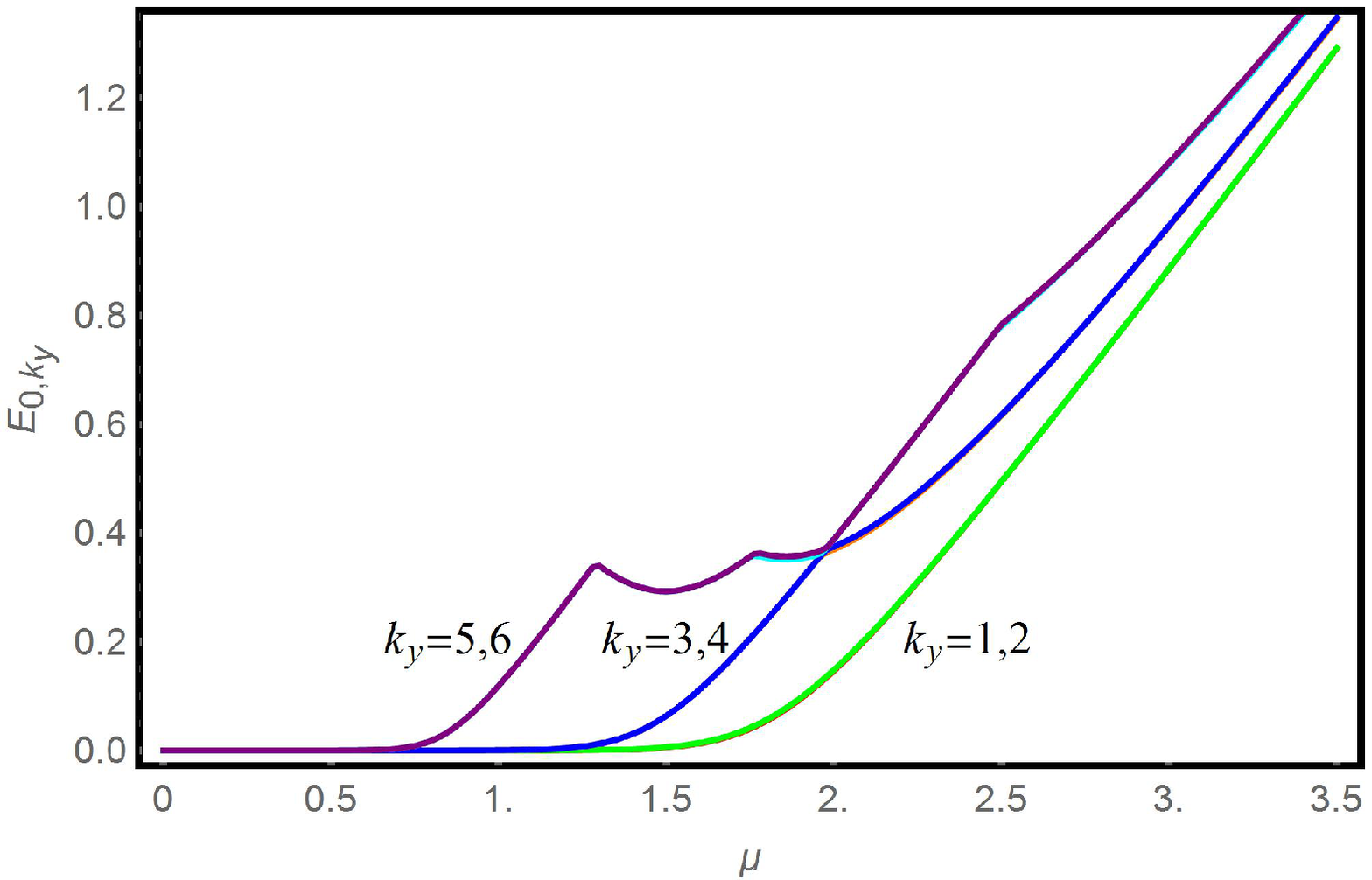}}
\caption{$E_{0,k_y}$ vs. chemical potential $\m$ for Kitaev ladder models with $N_x =120, N_y =6$ is shown, where $E_{0,k_y}$ are the energies of the six lowest positive energy modes.  $w_x=w_y=w$ is used as the unit of the energy. (a) top. For $w>|\D|$,(numerically $\D=0.8$ here), a pair of zero modes with a specific $k_y$ will exist when the corresponding condition in Eq.(\ref{zero_K2_2}) is satisfied.  The dashed lines indicates the values of $\m$, where various zero modes of the semi-infinite system begin to develop a gap. The numerical values are in good agreement with the analytic ones, and the discrepancy arises from the finiteness of $N_x$ used in the numerical calculation. (b) bottom. For $w<|\D|$, there is no zero mode even if $|\m|<4w.$ Here, we choose $\D=6.0$ so that the effect is more pronounced. }
\label{E0_vs_mu_Kitaev_2d_2}
\end{figure}

When both $N_x$ and $N_y$  become infinite, the condition for the existence of zero modes will be just $|\m_x|< 2w_x,$ and  $|\m_y| <2w_y$.  In sum, zero modes will exist when $|\m| < 2w_x + 2 w_y$.  

\subsection{IV. Chiral decomposition of superconducting nanowires with Dresselhaus spin-orbit coupling}
Now, let us apply similar analysis to superconducting nanowires with Dresselhaus spin-orbit coupling.  The corresponding Bogoliubov de-Gennes Hamiltonian is given by \cite{Ikegaya}
\bea
&\;& \hskip-0.6cm \HH_{2D}=\t_z\otimes\Id\otimes\left\{-\m \Id_x\otimes \Id_y -w \left[(\Mp)_x \otimes \Id_y + \Id_x \otimes (\Mp)_y \right] \right\} \nn \cr
&\;& \hskip 0.3cm - V_0 \t_3\otimes\s_1\otimes\Id_x\otimes \Id_y - \D_0 \t_2\otimes\s_2 \otimes\Id_x \otimes \Id_y  \nn \cr
&\;& \hskip 0.3cm  + \frac{i\lam_D}{2} \Id\otimes\s_3\otimes(\Mm)_x\otimes \Id_y.
\eea
Here, $\D_0$ is the $s$-wave pairing amplitude, $\lam_D$ the strength of Dresselhaus spin-orbit coupling, and $V_0$ the external Zeeman potential.  First of all, it is obvious that one can easily diagonalize the $y$-direction part, since only the term involving $(\Mp)_y$ is non-diagonal in that direction.  The boundary conditions at $j_y =0, N_y+1$ dictate that the eigenvalues of $(\Mp)_y$ is $2\cos p_y$, with $p_y=\pi k_y/(N_y+1), k_y = 1,\ldots, N_y.$  With this in mind, the above Hamiltonian may effectively be reduced to  one-dimensional  
\bea
&\;& \hskip-2.6cm \HH_{1D}=\t_z\otimes\Id\otimes\left\{-\m_{\rm eff}  \Id_x - w (\Mp)_x \right\} \nn \cr
&\;& \hskip -1.7cm - V_0 \t_3\otimes\s_1\otimes\Id_x - \D_0 \t_2\otimes\s_2 \otimes\Id_x \nn \cr 
&\;& \hskip -1.7cm+ \frac{i\lam_D}{2} \Id\otimes\s_3\otimes(\Mm)_x,
\eea
with $\m_{\rm eff} = \m + 2w \cos p_y.$
The explicit form of the chiral operator is now $\Pi = \t_2\otimes\s_1$, and the corresponding unitary transformation to bring $\HH_{1D}$ into the canonical form is
\be
\SS_2 = \frac{1}{\sqrt{2}} 
\left(\begin{matrix}
1 & 0 & 0 & i \cr
0 & 1 & i & 0 \cr
1 & 0 & 0 & -i \cr
0 & 1 & -i & 0 \cr
\end{matrix}\right) \otimes \Id_x.
\ee  
After the unitary transformation, the Hamiltonian takes the form
\bea
&\;& \tilde{\HH}_{1D} = \t_+\otimes\tilde{\HH}_{1D,{\rm R}} + \t_-\otimes\tilde{\HH}_{1D,{\rm L}}, \cr
&\;& \hskip -0.8cm {\rm with}\; \tilde{\HH}_{1D,{\rm R}} = \Id\otimes\left\{ \m_{\rm eff} \Id_x +w (\Mp)_x \right\} + V_0\, \s_1\otimes \Id_x  \nn\cr
&\;& \hskip 1.2cm - i \s_3 \otimes \left\{\D_0 \Id_x +\frac{\lam_D}{2} (\Mm)_x \right\}, \nn\cr
&\;& \tilde{\HH}_{1D,{\rm L}} = \Id\otimes\left\{ \m_{\rm eff} \Id_x +w (\Mp)_x \right\} + V_0\, \s_1\otimes \Id_x  \nn\cr
&\;& \hskip 1.2cm + i \s_3 \otimes \left\{\D_0 \Id_x +\frac{\lam_D}{2} (\Mm)_x \right\}. \nn
\eea
The Bogoliubov de-Gennes equation is now
\bea
&\;& \hskip -1.8cm \tilde{\HH}_{1D,{\rm R}} \tilde{\UU}_{\rm R} = E\tilde{\UU}_{\rm L}, \quad \tilde{\UU}_{1D,{\rm L}} \tilde{\UU}_{\rm L} = E\tilde{\UU}_{\rm R}.
\eea
Notice that here $\tilde{\UU}_{\rm R}$ and $\tilde{\UU}_{\rm L}$ still carry spin degree of freedom. 
Again, the left and right-handed zero modes decouples. In spin components, the right-handed part satisfies 
\bea
&\;& \hskip -2.0cm  \left\{ \m_{\rm eff}-i\D_0 +w (\Mp)_x - \frac{i\lam_D}{2} (\Mm)_x\right\} \tilde{\UU}_{{\rm R}\uparrow} \cr
&\;& \hskip -2.0cm + V_0 \tilde{\UU}_{{\rm R}\downarrow} =0, \cr
&\;& \hskip -2.0cm  \left\{ \m_{\rm eff}+i\D_0 +w (\Mp)_x + \frac{i\lam_D}{2} (\Mm)_x\right\} \tilde{\UU}_{{\rm R}\downarrow}\cr
&\;& \hskip -2.0cm + V_0 \tilde{\UU}_{{\rm R}\uparrow} =0. 
\eea
To solve the above equations, again we let $[\tilde{\UU}_{{\rm R}\uparrow}]_{j_x}=(\tilde{u}_0)_{{\rm R}\uparrow}s^{j_x}, [\tilde{\UU}_{{\rm R}\downarrow}]_{j_x}=(\tilde{u}_0)_{{\rm R}\downarrow}s^{j_x}.$  The above equations may be simplified to 
\bea
&\;& \hskip -1.8cm  \left\{ \m_{\rm eff}-i\D_0 + w( s+s^{-1} ) - \frac{i\lam_D}{2} (s-s^{-1}) \right\} (\tilde{u}_0)_{{\rm R}\uparrow} \cr
&\;& \hskip -1.7cm + V_0 (\tilde{u}_0)_{{\rm R}\downarrow}=0, \cr
&\;& \hskip -1.8cm  \left\{ \m_{\rm eff}+i\D_0 +w ( s+s^{-1} ) + \frac{i\lam_D}{2} (s-s^{-1}) \right\} (\tilde{u}_0)_{{\rm R}\downarrow} \cr
&\;& \hskip -1.7cm + V_0 (\tilde{u}_0)_{{\rm R}\uparrow}=0. \label{eqzeroD}
\eea
To have non-trivial solutions, we require the determinant of the two coefficients of the above linear equations to be zero. This gives rise to the following secular equation:
\bea
&\;& \hskip -0.5cm g(s)=\left[ \m_{\rm eff}+w(s+s^{-1}) \right]^2 + \left[ \D_0 + \frac{\lam_D}{2}(s-s^{-1}) \right]^2-V_0^2  \cr
&\;& \hskip 0.2cm =0. \label{PfD}
\eea
For simplicity, let's consider the infinite $N_x$ limit first. In such a case, the condition for the existence of zero modes may be easily determined by 
\bea
&\;& \hskip -2.3cm g(1)g(-1)= \left[ (\m_{\rm eff}+2w )^2 +(\D_0^2-V_0^2) \right] \cr
&\;& \hskip -0.6cm \times \left[ (\m_{\rm eff}- 2w)^2 +(\D_0^2-V_0^2) \right]<0.
\eea
Again, the product gives nothing but the Paffian of the zero modes.
After simplification, it is equivalent to 
\bea
&\;& \hskip -2.0cm {\rm For}\; |V_0|>|\D_0|\; {\rm and}\; 2w>\sqrt{V_0^2-\D_0^2} \cr
&\;& \hskip -2.0cm 2w - \sqrt{V_0^2-\D_0^2} < |\m_{\rm eff}| < 2w + \sqrt{V_0^2-\D_0^2}. \label{zero_DH_1}
\eea
\bea
&\;& \hskip -2.0cm {\rm For}\; |V_0|>|\D_0|\; {\rm and}\; \sqrt{V_0^2-\D_0^2}>2w,  \cr
&\;& \hskip -2.0cm \sqrt{V_0^2-\D_0^2} - 2w < |\m_{\rm eff}| < \sqrt{V_0^2-\D_0^2} + 2w. \label{zero_DH_2}
\eea
Substituting the expression for $\m_{\rm eff}$ into the above equations, we may determine the range of $\m$, within which a zero mode with a specific $k_y$ exists.  A numerical check for the above results is shown in Fig. 4, when $|V_0|>|\D_0|$.  It can be seen that the numerical values of the lower and upper critical values of $\m$ are in good agreement with the analytic ones.  We also show in Fig.5 that there is no zero mode when $|V_0|<|\D_0|.$
\begin{figure}[!h]
\centerline{\includegraphics[width=0.45\textwidth]{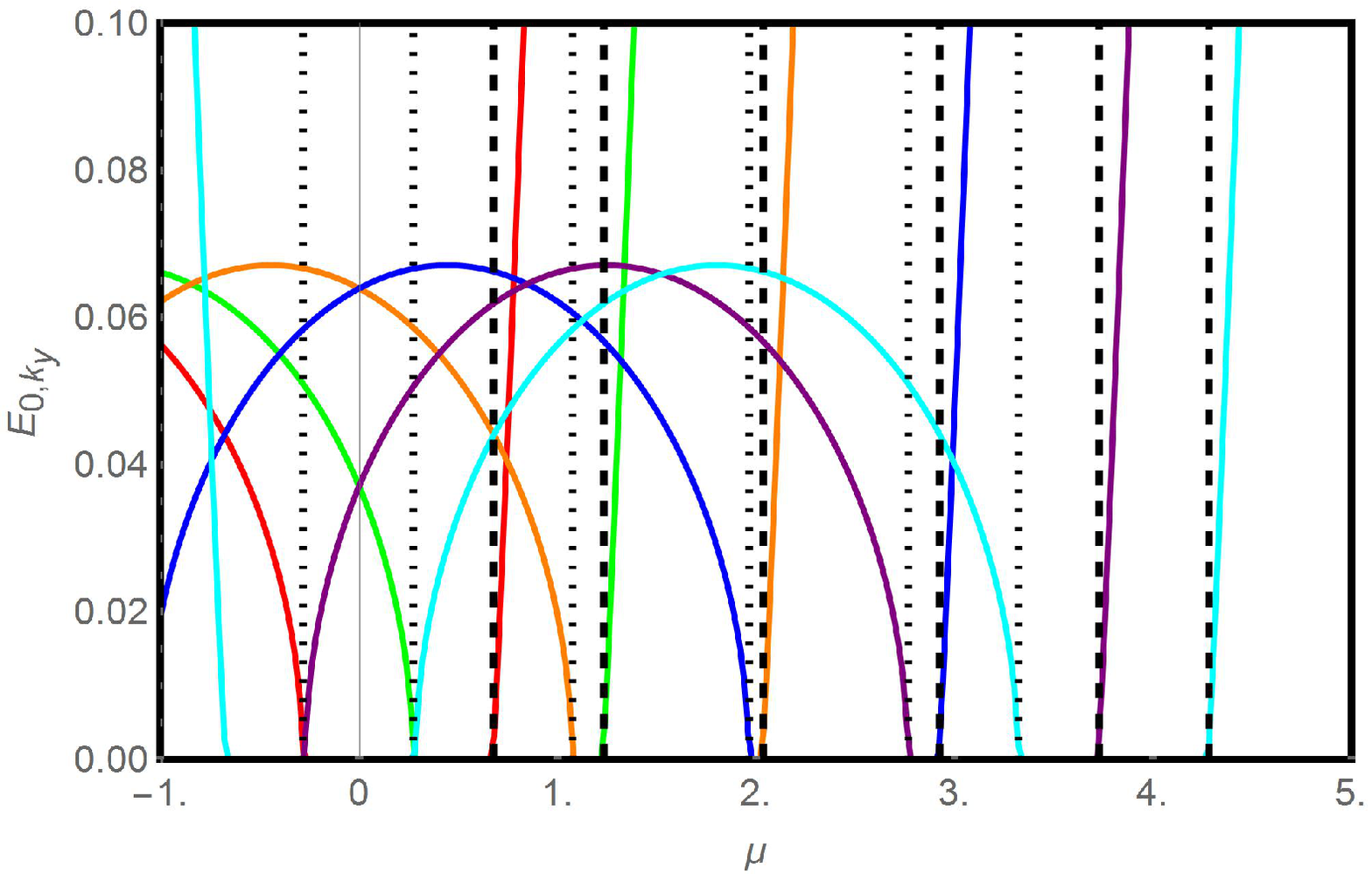}}
\centerline{\includegraphics[width=0.45\textwidth]{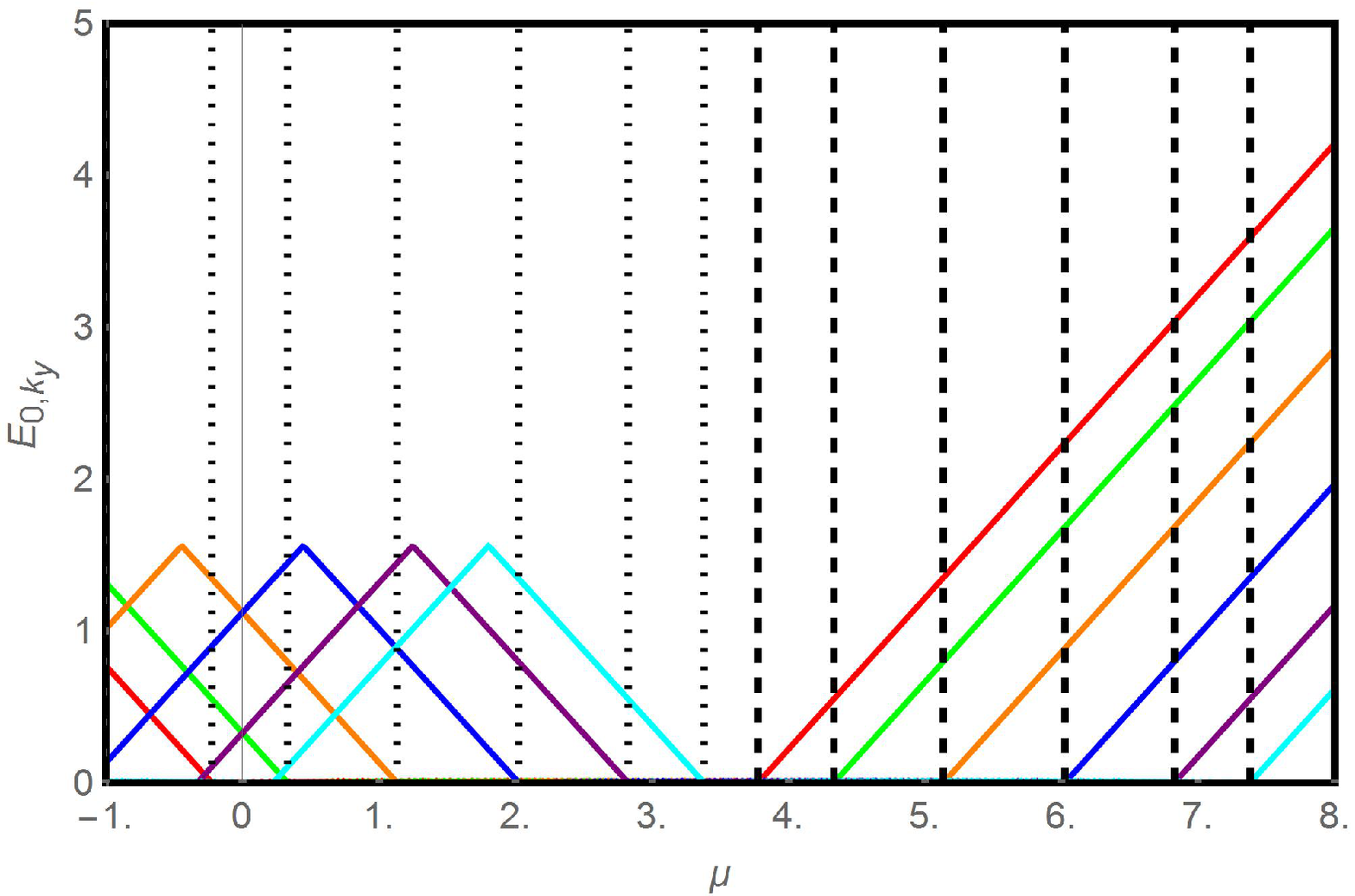}}
\caption{$E_{0,k_y}$ vs. chemical potential $\m$ for nanowires with Dresselhaus spin-orbit coupling is shown with $N_x =120, N_y =6$.  Here, $E_{0,k_y}$ are the energies of the six lowest positive energy modes, with $k_y=1-6$ from left to right. The dotted and dashed lines indicates the lower and upper critical values of $\m$ respectively, beyond which various zero modes of the semi-infinite system begin to develop a gap. (a) top. For $|V_0|>|\D_0|$, and $2w>\sqrt{V_0^2-\D_0^2}$, (numerically $V_0=0.8, \D_0=0.64$), a pair of zero modes with a specific $k_y$ will exist when the corresponding condition in Eq.(\ref{zero_DH_1}) is satisfied.   (b) bottom.  For $|V_0|>|\D_0|$, and $2w<\sqrt{V_0^2-\D_0^2}$, (numerically $V_0=3.6, \D_0=0.4$), a pair of zero modes with a specific $k_y$ will exist when the corresponding condition in Eq.(\ref{zero_DH_2}) is satisfied.  }
\label{E0_vs_mu_Dresselhaus_a}
\end{figure}

Alternatively, we may re-write the previous expressions as follows 
\bde
\item{1.} For $2w>\sqrt{V_0^2-\D_0^2}$ 
\bea
&\;& \hskip -0.5cm  \frac{2w -\m - \sqrt{V_0^2-\D_0^2}}{2w} < \cos p_y < \frac{2w -\m + \sqrt{V_0^2-\D_0^2}}{2w}, \cr
&\;& \hskip -0.5cm  {\rm or}\; \cr
&\;& \hskip -0.5cm \frac{-2w -\m - \sqrt{V_0^2-\D_0^2}}{2w} < \cos p_y < \frac{-2w -\m + \sqrt{V_0^2-\D_0^2}}{2w}. \cr
&\;& \hskip -0.5cm
\eea
\item{2.} For $2w<\sqrt{V_0^2-\D_0^2}$ 
\bea
&\;& \hskip -0.5cm \frac{\sqrt{V_0^2-\D_0^2} -\m -2w }{2w} < \cos p_y < \frac{\sqrt{V_0^2-\D_0^2} -\m +2w }{2w}, \cr
&\;& \hskip -0.5cm  {\rm or}\; \cr
&\;& \hskip -0.5cm \frac{-\sqrt{V_0^2-\D_0^2} -\m -2w }{2w} < \cos p_y < \frac{-\sqrt{V_0^2-\D_0^2} -\m +2w }{2w}. \cr
&\;& \hskip -0.5cm
\eea
\ede
This may be used to predict the number of zero modes for a given set of $\m, w, \lam_D, V_0, \D_0$ and $N_y$.

\begin{figure}[t]
\centerline{\includegraphics[width=0.45\textwidth]{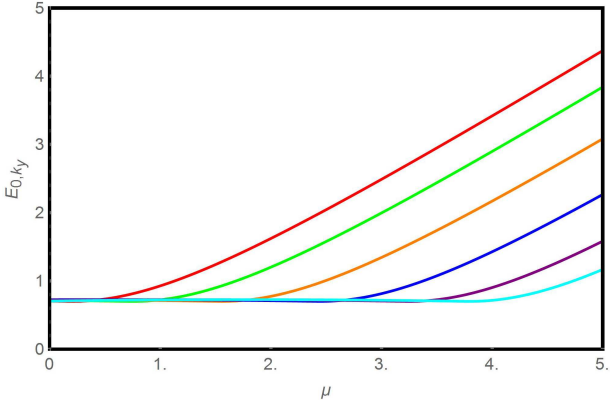}}
\caption{$E_{0,k_y}$ vs. chemical potential $\m$ for nanowires with Dresselhaus spin-orbit coupling with $N_x =120, N_y =6$ is shown.  Here, $E_{0,k_y}$ are the energies of the six lowest positive energy modes, with $k_y=1-6$ from left to right. For $|V_0|<|\D_0|$, there is no zero mode for any value of $\m.$ Numerically, we choose $V_0=0.6, \D_0=1.3$ here.}
\label{E0_vs_mu_Dresselhaus_b}
\end{figure}

The mathematical implication of the condition in Eq.(\ref{PfD}) is that the secular equation has either one or three real roots in the interval $[-1, 1]$. Since all the coefficients of the secular equation are real, complex roots always appear in conjugate pairs.  Accordingly, the secular equation has either two or four real roots. From Eq.(\ref{eqzeroD}), we know the spin down component will be determined by its corresponding spin up component:
\bea
&\;& \hskip -0.5cm (\tilde{u}_0^a)_{{\rm R}\downarrow} = \cr
&\;& \hskip -0.5cm  -\frac{(\tilde{u}_0^a)_{{\rm R}\uparrow}}{V_0} \biggl[\m_{\rm eff} - i\D_0 + w(s_a + s_a^{-1}) -i\frac{\lam_D}{2}(s_a - s_a^{-1}) \biggr]. \nn
\eea
As a result, we have four remaining free coefficients in the general solutions: 
\bea
\tilde{u}_{\rm R}(j_x)=\sum_{a=1}^{4}(\tilde{u}_0^a)_{\rm R} s_a^{j_x}.
\eea
The boundary condition at $j_x=0$ requires that 
\bea
\sum_{a=1}^{4}(\tilde{u}_0^a)_{\rm R} =0, \label{eqLD}
\eea
which further removes two more degrees of freedom. For the solution to be bounded, all the coefficients corresponding to \hfil\break
$|s_a|>1$ must be vanishing. If the number of roots with modulus larger than 1 is equal or larger than two, then there would not be any non-trivial solutions.  On the other hand, from the coefficients of the secular equation, we know the product of all the roots must be 1.  Therefore,  right-handed zero modes may exist only if three of the moduli are less than 1 and the other larger than 1. Accordingly, three of the moduli are larger than 1 and the other less than 1 for the left-handed zero modes.  As a result, there is no non-trivial solutions of the left-handed zero modes, as mentioned before.  

When $N_x$ is finite, the following condition must also be satisfied 
\bea
\sum_{a=1}^{4}(\tilde{u}_0^a)_{\rm R}s_a^{N_x +1} =0, \label{eqRD}
\eea
in addition to the condition in Eq.(\ref{eqLD}).  If the all the four roots to the secular equation are real, the boundary condition at $j_x =0, N_x+1$ can not both be satisfied.  Therefore, we only need to consider the case of two real and a conjugate pair of complex roots. In such a case, we may parameterize the roots as
\bea
&\;& \hskip -1.5cm s_1 =e^{-\b_1+i p_x},\; s_2 =e^{-\b_1-i p_x}, \cr 
&\;& \hskip -1.5cm s_3 =\pm e^{-\b_2},\; s_4 =\pm e^{2\b_1+\b_2},
\eea
with $\b_1, \b_2>0.$  Here, we have made use of the knowledge we gained from previous analysis: three of the roots has moduli less than 1 and the product of the four roots is 1.  To have non-trivial solutions, the determinant formed from the four free coefficients in Eq.(\ref{eqLD}) and Eq.(\ref{eqRD}) must be vanishing.  After simplification, the condition leads to
\bea
&\;& \hskip -1.5cm \b_2=\b_1 =\b,  \cr
&\;& \hskip -1.5cm p_x = \frac{2\pi k_x}{N_x+1},\; k_x =1,\ldots, \left[(N_x-1)/3 \right], 
\eea
where the square bracket stands for the Gauss symbol. 
By comparing with Eq.(\ref{PfD}), we may express $\cosh\b, \sinh\b$ and $\cosh(2\b)$ in terms of $\m_{\rm eff}, w, \D_0, V_0, \lam_D$ and $\cos(p_x)$:
\bea
&\;& \hskip -0.5cm \cosh\b = \cr
&\;& \hskip -0.5cm \frac{-4w \m_{\rm eff} (1+2\cos p_x)}{2\left[\m_{\rm eff}^2+4w^2-V_0^2+\D_0^2 \right]+ (4w^2+\lam_D^2)\left[\cos p_x + \cos(2p_x)\right]}, \cr
&\;& \cr 
&\;& \hskip -0.5cm \sinh\b = \cr
&\;& \hskip -0.5cm \frac{2\D_0 \lam_D (1+2\cos p_x)}{-2\left[\m_{\rm eff}^2-\lam_D^2-V_0^2+\D_0^2 \right]+ (4w^2+\lam_D^2)\left[\cos p_x + \cos(2p_x)\right]}, \cr
&\;& \hskip -0.5cm \cosh(2\b) = \frac{ 2\m_{\rm eff}^2+4w^2-\lam_D^2-2V_0^2+2\D_0^2 }{(4w^2+\lam_D^2)(1+2\cos p_x)}. \cr
&\;& \hskip -0.5cm 
\eea
Making use of the relations between $\cosh \b, \sinh \b$ and $\cosh(2\b)$, we can establish the following two identities involving $\m_{\rm eff}, w, \D_0, V_0, \lam_D$ and $\cos p_x$:
\bea
&\;& \hskip -0.5cm 4\left(4 w^2+ \lam_D^2 \right)^3 \cos^5p_x  \cr
&\;& \hskip -0.5cm + 4\left(4 w^2+ \lam_D^2 \right)^2 \left(\mef^2 +8 w^2 + \lam_D^2-V_0^2 +\D_0^2\right)\cos^4p_x \cr
&\;& \hskip -0.5cm + \left(4 w^2+ \lam_D^2 \right) \bigl(- 80 w^2\mef^2 + 12\lam_D^2 \mef^2 + 144 w^4 + 24 \lam_D^2 w^2 \cr 
&\;& \hskip -0.5cm - 48 V_0^2 w^2  + 48 \D_0^2 w^2 - 3\lam_D^4 - 12 V_0^2\lam_D^2 + 12 \D_0^2 \lam_D^2  \bigr) \cos^3p_x \cr 
&\;& \hskip -0.5cm + \left(4 w^2+ \lam_D^2 \right) \bigl(8 \mef^4 -124 w^2\mef^2 + \lam_D^2 \mef^2 - 16 V_0^2\mef^2  \cr
&\;& \hskip -0.5cm +16 \D_0^2 \mef^2 + 112 w^4 - 12\lam_D^2 w^2 - 68 V_0^2 w^2 + 68 \D_0^2 w^2 \cr
&\;& \hskip -0.5cm - 2\lam_D^4 - V_0^2\lam_D^2 + \D_0^2 \lam_D^2 + 8 V_0^4 - 16\D_0^2 V_0^2 + 8 \D_0^4 \bigr) \cos^2p_x \cr
&\;& \hskip -0.5cm + \left( 4 w^2+ \lam_D^2 \right) \bigl(8\mef^4 - 56 w^2\mef^2  - 6\lam_D^2 \mef^2 - 16 V_0^2\mef^2 \cr
&\;& \hskip -0.5cm + 16 \D_0^2 \mef^2 + 48 w^4 - 16 \lam_D^2 w^2 - 40 V_0^2 w^2  + 40 \D_0^2 w^2 + \lam_D^4 \cr
&\;& \hskip -0.5cm + 6V_0^2\lam_D^2 - 6\D_0^2 \lam_D^2 + 8 V_0^4 - 16 \D_0^2 V_0^2 + 8 \D_0^4  \bigr)\cos p_x \cr
&\;& \hskip -0.5cm + 4 \mef^6 + 32 w^2\mef^4 - 4 \lam_D^2 \mef^4 - 12 V_0^2\mef^4 + 12 \D_0^2 \mef^4 \cr
&\;& \hskip -0.5cm + 16 w^4\mef^2 - 40 \lam_D^2 w^2 \mef^2  - 64 V_0^2 w^2 \mef^2 + 64 \D_0^2 w^2\mef^2 \cr
&\;& \hskip -0.5cm + \lam_D^4 \mef^2 + 8 V_0^2\lam_D^2 \mef^2 - 8 \D_0^2 \lam_D^2 \mef^2 + 12 V_0^4\mef^2  \cr
&\;& \hskip -0.5cm - 24 \D_0^2 V_0^2 \mef^2 + 12 \D_0^4 \mef^2 + 64 w^6  - 32 \lam_D^2 w^4 -80 V_0^2 w^4  \cr
&\;& \hskip -0.5cm + 80\D_0^2 w^4 + 4\lam_D^4 w^2 + 24V_0^2 \lam_D^2  w^2 - 24 \D_0^2 \lam_D^2 w^2 + 32 V_0^4 w^2 \cr
&\;& \hskip -0.5cm - 64 \D_0^2 V_0^2 w^2 + 32 \D_0^4 w^2 - V_0^2 \lam_D^4 + \D_0^2 \lam_D^4 - 4V_0^4 \lam_D^2 \cr
&\;& \hskip -0.5cm + 8 \D_0^2 V_0^2 \lam_D^2 - 4 \D_0^4 \lam_D^2 - 4 V_0^6 + 12 \D_0^2 V_0^4 - 12\D_0^4 V_0^2 + 4 \D_0^6 \cr
&\;& \hskip -0.5cm =0. \label{zero_DH_3}
\eea
\vfill\eject
\bea
&\;& \hskip -0.5cm -4\left(4 w^2+ \lam_D^2 \right)^3 \cos^5p_x  \cr
&\;& \hskip -0.5cm + 4\left(4 w^2+ \lam_D^2 \right)^2 \left(\mef^2 - 4w^2 - 2\lam_D^2 - V_0^2 + \D_0^2\right)\cos^4p_x \cr
&\;& \hskip -0.5cm + \left(4 w^2+ \lam_D^2 \right) \bigl(48 w^2\mef^2 + 12\lam_D^2 \mef^2  + 48 w^4 - 24 \lam_D^2 w^2 \cr 
&\;& \hskip -0.5cm - 48 V_0^2 w^2  + 48 \D_0^2 w^2 - 9\lam_D^4 - 12 V_0^2\lam_D^2 - 20 \D_0^2 \lam_D^2  \bigr) \cos^3p_x \cr 
&\;& \hskip -0.5cm + \left(4 w^2+ \lam_D^2 \right) \bigl(-8 \mef^4 + 4 w^2\mef^2 + 17\lam_D^2 \mef^2 + 16 V_0^2\mef^2  \cr
&\;& \hskip -0.5cm - 16 \D_0^2 \mef^2 + 32 w^4 + 12\lam_D^2 w^2 - 4 V_0^2 w^2 + 4 \D_0^2 w^2 - 7\lam_D^4 \cr
&\;& \hskip -0.5cm - 17V_0^2\lam_D^2 - 31\D_0^2 \lam_D^2 - 8 V_0^4 + 16\D_0^2 V_0^2 - 8 \D_0^4 \bigr) \cos^2p_x \cr
&\;& \hskip -0.5cm - \left( 4 w^2+ \lam_D^2 \right) \bigl(8\mef^4 + 24 w^2\mef^2 - 10\lam_D^2 \mef^2 - 16 V_0^2\mef^2 \cr
&\;& \hskip -0.5cm + 16 \D_0^2 \mef^2 + 16 w^4 - 16 \lam_D^2 w^2 - 24 V_0^2 w^2  + 24 \D_0^2 w^2 + 3\lam_D^4 \cr
&\;& \hskip -0.5cm + 10 V_0^2\lam_D^2 + 14 \D_0^2 \lam_D^2 + 8 V_0^4 - 16 \D_0^2 V_0^2 + 8 \D_0^4  \bigr)\cos p_x \cr
&\;& \hskip -0.5cm + 4 \mef^6 + 16 w^2\mef^4 - 8 \lam_D^2 \mef^4 - 12 V_0^2\mef^4 + 12 \D_0^2 \mef^4 \cr
&\;& \hskip -0.5cm + 16 w^4\mef^2 - 24 \lam_D^2 w^2 \mef^2 - 32 V_0^2 w^2 \mef^2 + 32 \D_0^2 w^2\mef^2 \cr
&\;& \hskip -0.5cm + 5\lam_D^4 \mef^2 + 16 V_0^2\lam_D^2 \mef^2 - 16 \D_0^2 \lam_D^2 \mef^2 + 12 V_0^4\mef^2  \cr
&\;& \hskip -0.5cm - 24 \D_0^2 V_0^2 \mef^2 + 12 \D_0^4 \mef^2 - 16 \lam_D^2 w^4 - 16 V_0^2 w^4 + 16\D_0^2 w^4  \cr
&\;& \hskip -0.5cm + 8\lam_D^4 w^2 + 24V_0^2 \lam_D^2 w^2 - 40 \D_0^2 \lam_D^2 w^2 + 16 V_0^4 w^2 \cr
&\;& \hskip -0.5cm - 32 \D_0^2 V_0^2 w^2 + 16 \D_0^4 w^2 - \lam_D^6 - 5V_0^2 \lam_D^4 + \D_0^2 \lam_D^4 - 8V_0^4 \lam_D^2  \cr
&\;& \hskip -0.5cm + 16 \D_0^2 V_0^2 \lam_D^2 - 8 \D_0^4 \lam_D^2 - 4 V_0^6 + 12 \D_0^2 V_0^4 - 12\D_0^4 V_0^2 + 4 \D_0^6 \cr
&\;& \hskip -0.5cm = 0. \label{zero_DH_4}
\eea

From these identities, we would be able to numerically find the region in the parameter space where exact zero modes exist for finite $N_x$. Since the above expressions are quite tedious, here we will only give a numerical example in Fig.6 to show such zero modes do exist without going into detailed investigation.
\begin{figure}[t]
\centerline{\includegraphics[width=0.45\textwidth]{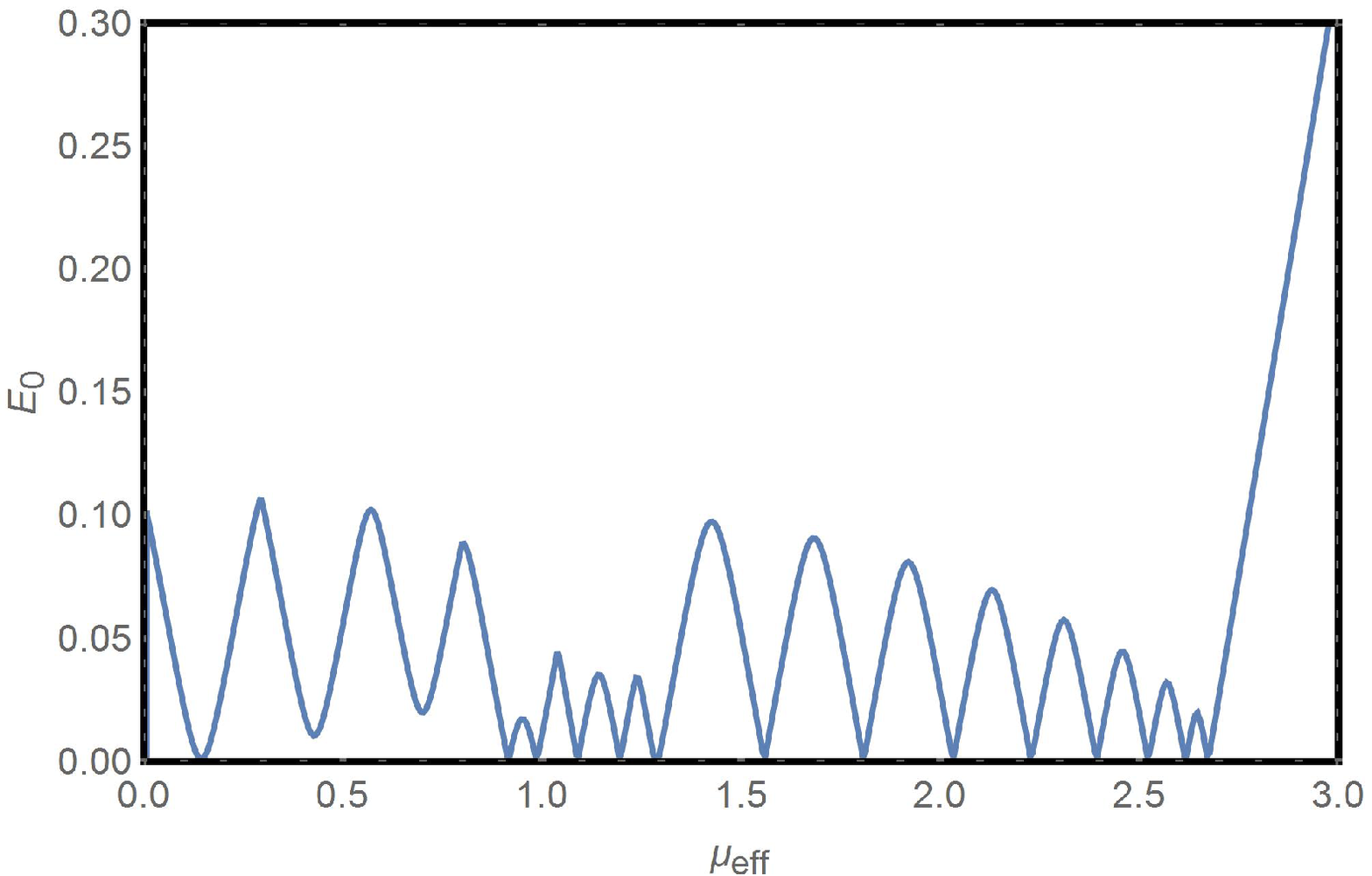}}
\centerline{\includegraphics[width=0.45\textwidth]{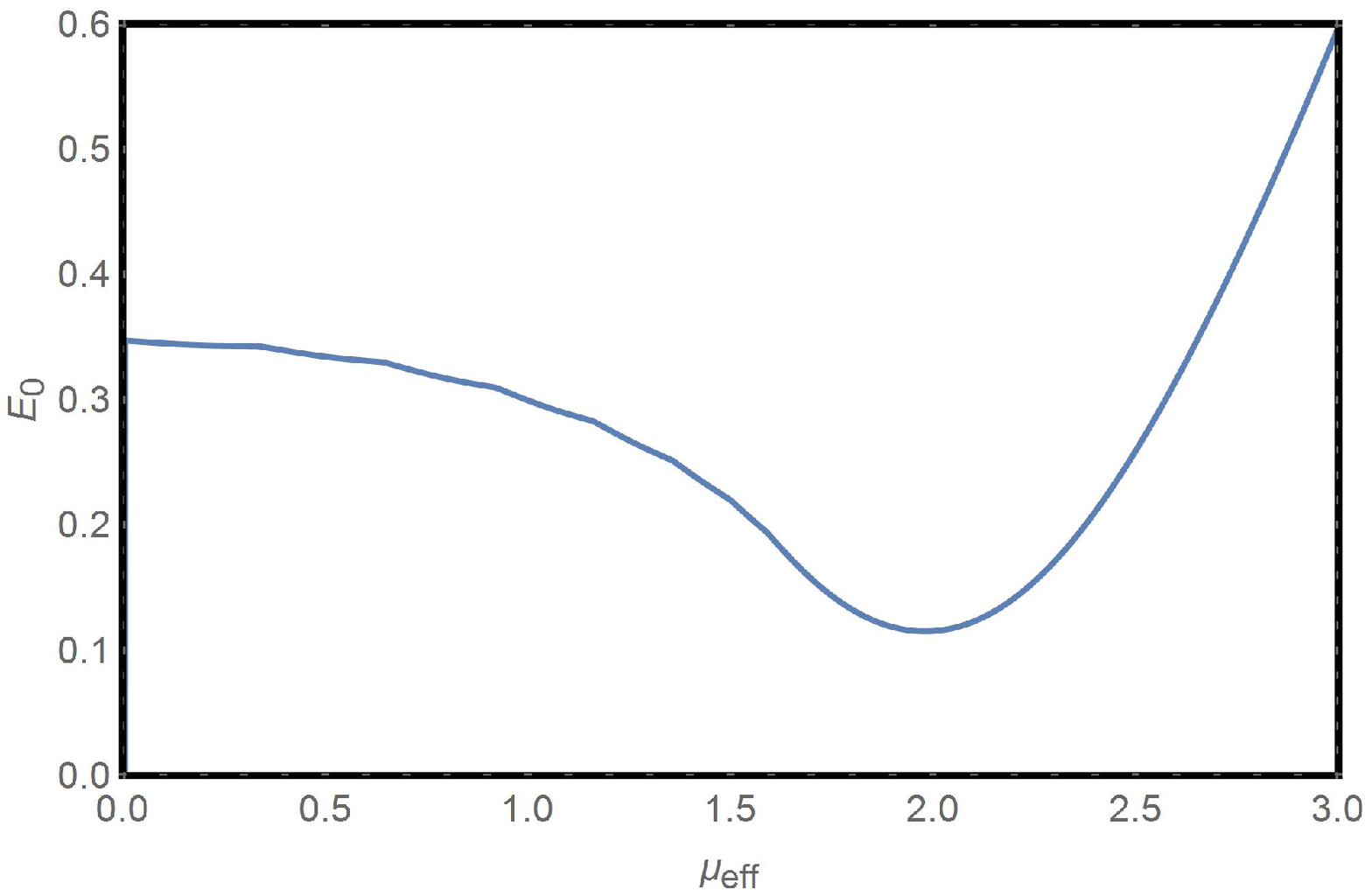}}
\caption{$E_0$ vs. chemical potential $\mef$ for nanowires with Dresselhaus spin-orbit coupling is shown, where $E_0$ is the energy of the lowest positive energy mode. Here, we choose $\lam_D= 0.5, V_0= 0.7,$ and $N_x=20$ (a) top. For $|V_0|>|\D_0|$ (numerically, $\D_0=0.1$), there are a pair of exact zero modes when the condition in Eqs.(\ref{zero_DH_3}) and (\ref{zero_DH_4}) are satisfied. The value of the exact zero mode energy is zero up to numerical inaccuracy, which is of the order $10^{-16}.$ (b) bottom.  For comparison, when $|V_0|<|\D_0|$ (numerically $\D_0=0.8$), there is no zero modes. }
\label{E0_vs_mueff_Dresselhaus}
\end{figure}

\subsection{V. Conclusion and discussion}

Using the chiral decomposition, we are able to find analytically the zero modes and the conditions for such modes to exist in the Kitaev ladder model and superconducting nanowires with Dresselhaus spin-orbit coupling.  Taking advantage of the explicit expressions of these conditions, we can calculate the number of zero modes of for the Kitaev ladder model for arbitrary parameters, and for the superconducting nanowires with Dresselhaus spin-orbit coupling in the infinite $N_x$ limit. Since only one of the left or right-handed zero modes survive in the limit, chiral symmetry of the zero modes is broken. Moreover, when suitable resonance condition is satisfied we find that exact zero modes may exist in these systems even if it is finite, contrary to what is usually believed.  The method of chiral decomposition used here is quite versatile.  Thus, it may also be applied to other setups, such as NS, SNS junctions, or the continuous limit of these systems as long as the existence of a chiral operator is intact.  In particular, we expect the analysis would simpler in the continuous limit, and the results should be qualitatively similar to what we have obtained here \cite{Ikegaya}.  Finally, we would like to mention that the two models we analyze here both belong to the BDI class, where the number of MZES may take on any integer values. According to the classification of topological insulators, there are many other systems with a chiral operator or other similar operator which anti-commute with the Hamiltonian of the system \cite{Ryu}. It would be interesting to carry out similar analysis to these systems.  By applying chiral decomposition to study the MZES in these systems, we may be able to obtain some analytic results which would enhance our understanding of their topologically non-trivial phases.

\section*{Acknowledgments}
The work is supported in part by the Grants 101-2112-M-003-002-MY3 of the National Science Council, Taiwan.

\end{document}